# Coupled Micro-Doppler Signatures of Closely Located Targets

Vitali Kozlov[1, *], Sergey Kosulnikov[1], Dmitry. Filonov[2], Andrey. Schmidt[1] and Pavel Ginzburg[1]

[1]School of Electrical Engineering, Tel Aviv University, Tel Aviv, 69978, Israel

[2]Center for Photonics and 2D Materials, Moscow Institute of Physics and Technology, Dolgoprudny, 141700 Russia

**Abstract:**

The classical Doppler shift originates from the movement of a target's center of mass, but it does not hold information about the internal dynamics of the scattering object. In contrast, micro-Doppler signatures contain data about the micro-motions that arise from internal degrees of freedom within the target (such as rotation and vibration), which can be remotely detected by careful analysis of the scattered field. Here we investigate, both theoretically and experimentally, how coupling between a pair of closely situated targets affects the resulting micro-Doppler signatures. The presented model considers a pair of near-field coupled resonators with dynamically reconfigurable scattering properties. Voltage controlled varactor diodes enable modulating the scattering cross-section of each target independently, mimicking rotational degrees of freedom. As a result, coupled micro-Doppler combs are observed, containing new frequency components that arise from the near field coupling, making it possible to extract information about the internal geometry of the system from far-field measurements. From a practical point of view, micro-Doppler spectroscopy allows remote classification of distant objects, while deep understanding of the coupling effects on such signatures in the low frequency regime can provide valuable insight for radar and sonar systems, as well as optical and stellar radio-interferometry, among many others.

*vitaliko@mail.tau.ac.il

**Introduction**

Remote sensing and identification of distant objects is important in many areas of fundamental and applied science. The length-scales involved in the observation can span from several microns in classical microscopy to light-years in stellar radiometry [1], where the most valuable information is the location of objects and the velocity of their centers of mass. While the focus of this investigation is concentrated around radar applications [2,3,4], the implications for related fields will be addressed in the concluding section.

Although the extraction of range and Doppler information is the main goal of radar systems, the internal degrees of freedom within the target (such as vibrations and rotations) create additional modulation, termed micro-Doppler shifts, which are useful for target classification [5]. Specifically, these can be rotating blades of airborne engines, including those of airplanes, helicopters, and drones. Furthermore, pedestrians, cars, cyclists and other types of moving objects have well defined and specific features in the reflected signals' spectra [6–14]. In principle, each type of target has a specific micro-Doppler signature that can be used for its remote identification and classification, as long as signal to noise ratio allows performing reliable analysis. Micro-Doppler processing strongly depends on the type of transmitted radar signal. In the short pulsed regime, the available information for processing is typically the phase accumulation between consecutive pulses [15], whereas for continuous wave systems the direct measurement of the spectrum, termed 'micro-Doppler comb', is more straightforward [16,17].

In order to obtain high range and azimuth resolution, most radar systems strive to operate in the high frequency range, well above 20 GHz [18–20]. For such high frequencies the analysis of the micro-Doppler signatures can be substantially simplified by subdividing the target into a grid of cells and employing ray tracing methods that assume no coupling between different mesh cells (i.e. no multi scattering events) [5,15,21]. The coupling effects in the low frequency regime, where the overall size of a target is comparable or even smaller than the impinging wavelength, are rarely studied in the literature. Recent work was done to shed light on the phenomena of long wavelength micro-Doppler signatures, revealing pronounced differences between the regimes [16,17]. Specifically, it was shown that rotating wires produce a discrete micro-Doppler comb, originating from the strong electric coupling between different points on the wire. Here we study the effect of near-field coupling on the micro-Doppler comb by analyzing a closely related model problem. Two split ring resonators (SRRs) with varactor diodes soldered in their gaps are placed in the near-field of each other, representing a system with two independent degrees of freedom. The resonant frequency of each SRR is modulated in time by controlling the voltage drop on its varactor diode. The result of this modulation is a micro-Doppler comb, which is similar to the one obtained from a single rotating wire. Different modulation frequencies are applied to the resonators, mimicking different motions within a complex target. Coupling between SRRs generates parametric mixing between the micro-Doppler combs, producing new frequencies in the spectrum of the scattered field. Since the applied time-modulation is orders of magnitude slower than the carrier frequency, a time scale separation approach is employed in the analysis. The coupled micro-Doppler combs are shown to be strongly dependent on the geometrical arrangement of the system.

The manuscript is organized as follows: the relation between parametric time-modulation of a resonator and micro-Doppler comb is presented first. It is followed by the time-varying coupled dipoles theoretical model, which is experimentally verified next. Extended discussion on the relevance of the effect to other physical systems comes before the conclusion.

**Results**

*Emulation of rotating wire's micro-Doppler comb by modulation of a dipoles resonant frequency*

Micro-Doppler frequency combs emerge when periodic modulation is applied to the scattering cross-sections of objects, subject to electromagnetic illumination [16]. A typical scenario here is the jet engine modulation of a radar signals [6,7]. It should be noted that the periodicity of the modulation needs to be much shorter than the illuminating pulse width, containing at least a few oscillations of the carrier. In this case, a sufficient number of modulation cycles will be accomplished in time during the pulse duration, producing a comb in frequency (Fourier transform of the time-dependent scattered signal). The frequency comb created by this internal motion can be mimicked by applying an electronic time-modulation to an antenna's resonant frequency. For example, consider a magnetic dipole (a simplified electromagnetic description of a split ring resonator) with time modulated polarizability

$$\alpha(t) = \frac{q}{\widetilde{\omega}_0^2(t) - \omega^2 + j\gamma\omega}, \quad (1)$$

where $q$ is the oscillator strength and $\gamma$ is related to the losses that determine the resonance width. The structure is immersed in an incident plane wave with the magnetic field $H_{inc} = H_0\, e^{jkx}\, e^{j\omega t}\hat{z}$, where $\omega$ is the carrier frequency and $k$ is the free-space wave number (recall that the propagation direction is in the negative x direction). The resonant frequency of the structure is given by an equivalent LC circuit model [22,23,24,25] as:

$$\widetilde{\omega_0}(t) = \frac{1}{\sqrt{LC(t)}} = \frac{\omega_0}{\sqrt{(1+\beta \sin(\Omega t))}}, \quad (2)$$

where L is the inductance of the circuit, C the capacitance and $\omega_0 = \frac{1}{\sqrt{LC(t=0)}}$. The capacitive load can be harmonically modulated in time with frequency $\Omega$ by using a varactor diode, which changes its nominal in response to an applied external voltage (amplitude $\beta$ is trivially related to the modulation amplitude). Assuming that the capacitance modulation is slow ($\frac{\Omega}{\omega} \ll 1$), a time-scale separation approach may be employed. Technically, it means that $sin(\Omega t)$ is considered a parameter within the linear set of frequency domain Maxwell's Equations. In this case, scattering from a subwavelength SRR is approximated by a point magnetic dipole with a certain polarizability. The parametric time-dependence of the resonance frequency, given by Eq. 2, produces a scattered field at the observation point along the X-axis (see the inset to Fig. 4). The complex scattering amplitude is proportional to the polarizability given by Eq. 1. Transforming the slow-time dependent scattered field into the frequency domain produces a micro-Doppler comb which is similar to that of a rotating wire [16], as shown in Fig. 1.

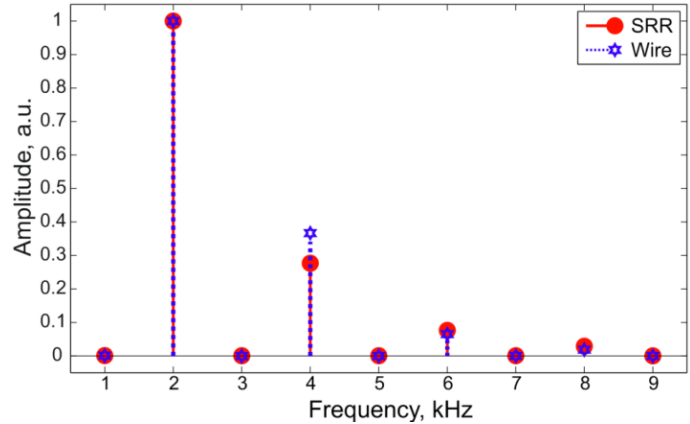

Figure 1: Theoretical micro-Doppler combs of a harmonically time-modulated magnetic dipole (a simplified electromagnetic model of a split ring resonator) and a rotating wire. Parameters of the systems are given in the main text.

In order to perform the comparison, both the wire and the SRR calculations were performed for incident plane wave excitation with 2.88 GHz carrier frequency. The wire's length is chosen to be 16 cm, while for the magnetic dipole $20\gamma = \omega_0 = 2.88 \times 2\pi\, GHz$, $\beta = 0.03$ and $q = 3 \times 10^{14}$ were chosen (the coefficients were obtained by fitting the model to numerical finite elements calculation). The wire was assumed to rotate at a frequency of 1 kHz, while the magnetic dipole was modulated at $\Omega = 2 \times 2\pi\, kHz$, insuring that both combs produce the maximum peak at 2 kHz. Both results were normalized to the intensity of the second peak in order to show the resemblance of the frequency combs, as appears on Fig. 1.

Rotational speeds of 1 kHz are quite typical for fast jet engines, which can even surpass these values. On the other hand, obtaining electronic modulation at those frequencies is rather trivial and straightforward task. Similarity between micro-Doppler combs, produced by direct time-modulation of the dipole's resonance and the rotating wire, suggest emulating mechanical system with an electronic one. Note that odd micro-Doppler peaks are absent in both of the idealized systems owing to the assumed perfectly symmetrized and balanced axial rotation of the wire, as well as small modulation amplitude of the magnetic dipole. As a remark, relevant to subsequent experimental realizations, it should be noted that the nonlinearity of the varactor diode can complicate the time dependence (the capacitance will also depend on the modulation amplitude, which will result in $\beta = \beta(t)$). Here the implication of this behavior is a distortion of the micro-Doppler peaks' amplitudes and the appearance of odd frequencies in the comb.

Time-dependent coupling between two point targets and the resulting coupled micro-Doppler peaks will be analyzed next.

*Parametrically time-dependent coupled dipoles model*

The coupled dipoles method is a well developed tool for solution of electromagnetic problems, e.g. [26], where mutual interactions between polarizable particles are taken into account via retarded Green's functions. Here this method is used to formulate matrix equations for calculating the backscattered field, e.g. [27]. The formulation is significantly simplified in the case of two dipoles with magnetic moments $m_1$ and $m_2$. Here, without a loss of generality, consider a case when the dipoles are situated in the Z-Y plane with polarizabilities in the z direction only (see the inset to Fig. 4), satisfying the self-consistent equation:

$$m_i = \alpha_i \left( \sum_{i \neq j} A_{ij} m_j + H_i^0 \right), \quad (3)$$

where $\alpha_i$ is the magnetic polarizability of particle '$i$', $H_i^0$ is the incident magnetic field at the location of particle '$i$' and $A_{ij} = A_{ji} \equiv A$ is the associated scalar Green's function connecting the dipole source in the location of particle '$j$' to its scattered field in the location of particle '$i$'. The polarizability is assumed to be of the Lorentzian shape, given by Eq. 1. Similar notations were used in [27,28], which contains the detailed mathematical formulation. While both particles are assumed to possess the same polarizability, they are modulated with different frequencies $\Omega_1$ and $\Omega_2$. Formula 3 is a set of two linear scalar equations, which can be self-consistently solved for the dipole moments. Assuming that the transmitting and receiving antennas are both polarized in the Z direction and are placed far apart from the dipoles, the solution for the scattered magnetic field at a point $L$ along the X-axis (backscattered field), polarized in the Z direction is given by:

$$H_L^s = \frac{\alpha_1(t) A_{L1} H_1^0 + \alpha_2(t) A_{L2} H_2^0 + \alpha_1(t)\alpha_2(t) A (A_{L1} H_2^0 + A_{L2} H_1^0)}{1 - A^2 \alpha_1(t)\alpha_2(t)}, \quad (4)$$

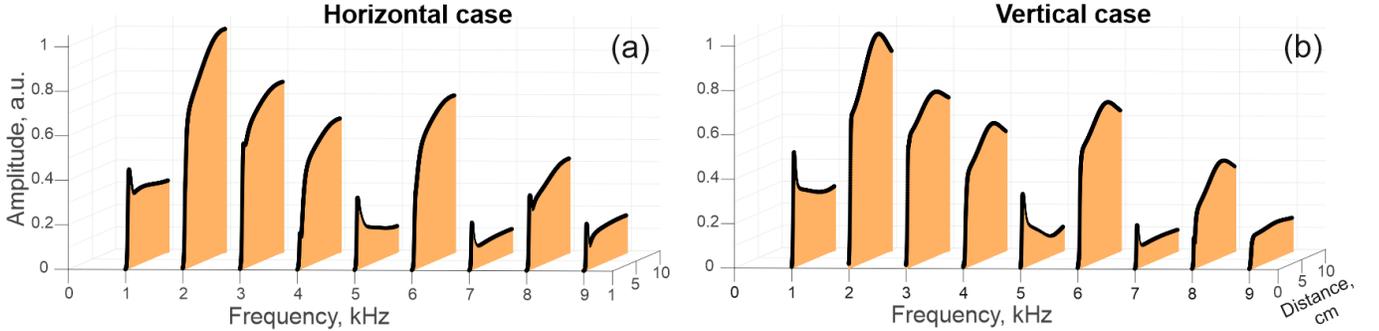

Figure 2: Theoretical plots of the micro-Doppler comb as a function of distance between the two modulated dipoles (produced using Eq. 4). (a) Horizontal configuration, where both dipoles are on the Y-axis. (b) Vertical configuration, where both dipoles are on the Z-axis. Parameters for both of the scenarios are provided in the main text.

where $A_{L1}$ and $A_{L2}$ are the associated scalar Green's function connecting the dipole moment '$i$' with its scattered magnetic field at a point $L$, located on the X-axis (see the inset to Fig. 4). It can be seen that $A$ determines the strength of the coupling. For $A = 0$ there is no coupling and the scattered field is simply a superposition of the two independent dipoles. Since typically $|A^2 \alpha_1(t)\alpha_2(t)| \ll 1$, the denominator of Eq.4 can be expanded as:

$$H_L^s(t) = [\alpha_1(t) A_{L1} H_1^0 + \alpha_2(t) A_{L2} H_2^0 + \alpha_1(t)\alpha_2(t) A (A_{L1} H_2^0 + A_{L2} H_1^0)] \sum_{n=0}^{\infty} (A^2 \alpha_1(t)\alpha_2(t))^n. \quad (5)$$

If the illumination is at the resonant frequency $\omega = \omega_0$ and the amplitude of the modulation $\beta$ is small, the polarizability in Eq. 1 may be approximated as,

$$\alpha_i(t) \approx \alpha_r \beta_i \sin(\Omega_i t) - j\alpha_I(1 - \beta_i^2 \sin^2(\Omega_i t)), \qquad (6)$$

where $\alpha_r$ is the slope of the real part of the polarizability near the resonance and $\alpha_I = q/\gamma\omega$ the peak of its imaginary part. Substituting Eq. 6 into Eq. 5 reveals a time harmonic series of all possible combinations of frequencies $\Omega_1$ and $\Omega_2$, including odd harmonics, as shown in Fig. 2 for different geometric configuration of the dipoles. The first configuration is when the two dipoles are located one above the other at a distance $d$ along the Z axis (vertical case) and the second is when the two dipoles are located at a distance $d$ on the Y axis (horizontal case). The dipoles are modulated with frequencies $\Omega_1 = 2 \times 2\pi\ kHz$, $\Omega_2 = 3 \times 2\pi\ kHz$ and amplitudes $\beta_1 = \beta_2 = 0.1$. The illumination frequency is chosen as $\omega = \omega_0 = 10\gamma = 2\pi \times 2.88\ [GHz]$, closely resembling the experimental conditions.

It can be observed that in both the horizontal and vertical configurations the Doppler comb consists of multiple harmonics, including the harmonics produced separately by each dipole (dipole 1: 2nd, 4th, 6th, and 8th harmonics; dipole 2: 3rd, 6th, 9th), as well as harmonics that did not appear in the single dipole case (1st, 5th and 7th harmonics). These additional harmonics are the result of coupling between the dipoles, and so they are more sensitive to the distance between the scatterers. The most representative one is the 1 KHz harmonic, which cannot be generated by either of the dipoles, but results from difference frequency generation (e.g. 3 KHz – 2 KHz and from other linear combinations, providing smaller contributions). It can be also seen that the amplitudes of the natural (unmixed) harmonics are stronger (e.g. 2nd and 3rd). Furthermore, 6th harmonics is generated by both of the SRRs in the uncoupled case and hence, its amplitude is relatively stronger than the surrounding contributions. The scalar Green's function for both configurations is proportional to $A \propto \frac{e^{ikd}}{d}$, which means the coupled peaks are expected to decay (and eventually oscillate) with distance, as seen on Fig 2 and 3 ((1st, 5th and 7th harmonics)). An additional observation could be made about the decay of all peaks as the distance between the dipoles tends to 0, corresponding to Eq.4, where the denominator diverges. It should be noted that at close distances the model is no longer accurate as the quadruple and higher moments become more dominant.

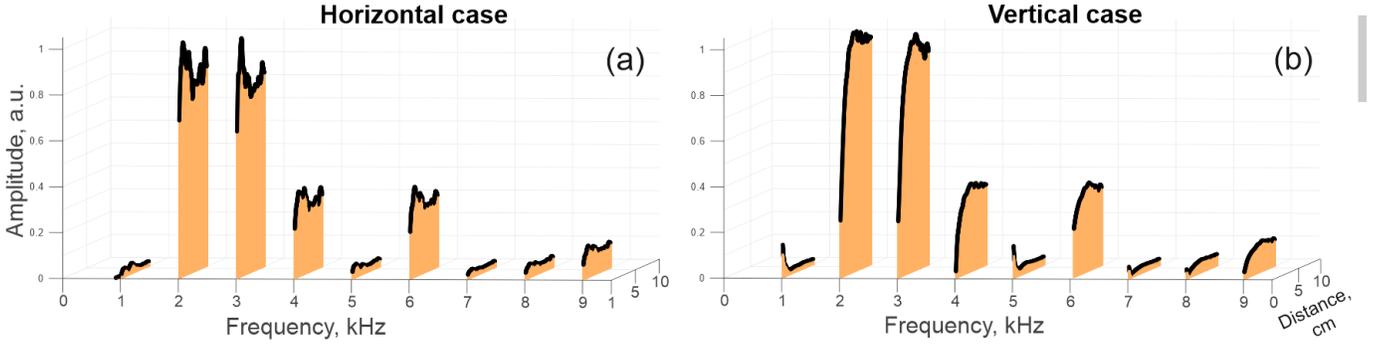

Figure 3: Full wave simulation of the frequency comb produced by two modulated split ring resonators as a function of the distance between them. (a) Horizontal configuration, where both SRRs are on the Y-axis. (b) Vertical configuration, where both SRRs are on the Z-axis. Parameters for both of the scenarios are provided in the main text.

*Micro-Doppler combs mixing - numerical model*

In order to corroborate these theoretical results, a full wave numerical analysis was performed. The simulation was done in multiple steps, choosing the instantaneous capacitance at the gap of each SRR, solving the scattering problem and then repeating the process for a different value of capacitance (representing the next step in the slow time). Each possible instantaneous combination of the capacitances was analyzed and then stitched together to form a true time sequence. This routine relies on the assumption of the slow time-scale parametric process, since standard finite element modeling cannot directly address time-varying problems. Additionally, the simulations were repeated for varying distances between the SRRs in both configurations to reveal the influence of the coupling effects. The numerical model of the structure was designed with CST Studio Suite, using the values provided in the experimental realization discussed ahead. The minimal distance between the rings' boundaries is 5mm in the vertical case and 1cm in the horizontal case (since the rings cannot be brought any closer in this configuration without overlapping).

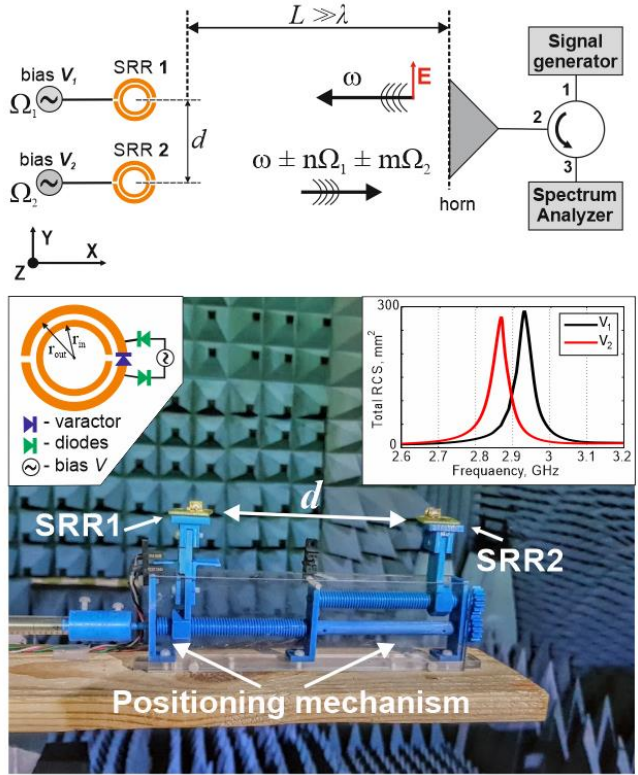

Figure 4: The experimental set up. Two split rings are mounted on a positioning mechanism, which allows controlling the distance between them. The setup is illuminated with a 2.88 GHz plane wave, polarized in the direction of magnetic polarizability of the rings. The resonant frequency of each SRR is separately modulated by applying a biasing voltage to the varactor diode placed in the gap. Left inset – schematic representation of the SRRs geometry and biasing network. Top inset – schematic representation of the experimental set up for the horizontal configuration. Right inset – the resonance shift caused by applying two different voltages ($V_1 = 3.5V$, $V_2 = 4.5V$).

A qualitative agreement between Fig. 2 and Fig. 3 can be observed, where the differences are attributed to the fact full wave simulations take into account all of the multipole expansion, whereas the dipole approximation does not contain contributions from higher orders. Those contributions can be further suppressed by employing more complex SRR's geometries for eliminating spurious electric dipole contributions and anisotropy [29]. For the Horizontal case, the theoretical plot starts from a distance of 0mm, where as the numerical model has a finite sized SRR, forcing the distance between the resonators to start at 1cm, where coupling is already weak and most of the decay of the 1st, 5th and 7th peaks already occurred, as well as the rise of the amplitudes of the 2nd, 3rd, 4th and 5th harmonics. In both theoretical and numerical investigations, the primary harmonics (2nd and 3rd) are the strongest and the 6th one, which is the their multiple, is also very pronounced. The emergence of the 1st peak is the clear indication of the coupling.

*Experimental Results*

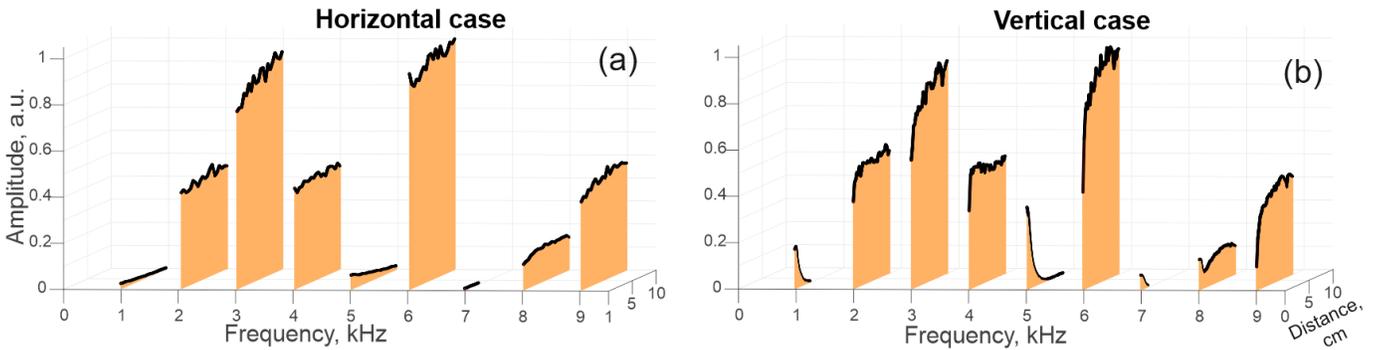

Figure 5: Experimental results of the micro-Doppler comb between two modulated SRRs as a function of distance. (a) Horizontal configuration, where both SRRs are on the Y-axis. (b) Vertical configuration, where both SRRs are on the Z-axis.

In order to test the theoretical predictions, two split rings were manufactured. In order to reduce the effect of the biasing network on the SRR resonance, two diodes were placed in series with the biasing source as shown in Fig. 4 (left inset).

The SRRs were mounted on a platform that allowed controlling the distance between them. A step motor, connected to a calibrated screw system, was used to precisely position the SRRs with respect to each other. Both sample holders and screws were implemented with a 3D printing technique, where PLA materials have been used (blue-colored constructions in Fig. 4). Typical permeability of those constructions are around 3 [30], which does not significantly affect the scattering from the SRRs. This accurate and repeatable positioning was implemented to achieve high experimental precision, which is necessary due to the sensitivity of the coupled micro-Doppler combs to variations in the geometry.

The magnetic dipoles were realized as a loaded double split ring resonator (DSRR) with inner ring radius $R_{in} = 3$ mm, width of the inner ring radius $W_{in} = 0.5$ mm, outer ring radius $R_{out} = 4.25$ mm, outer ring width $W_{out} = 1$ mm and two gaps (with no load for the inner ring and with loaded outer one) $G_{gap} = 0.5$ mm (see the left inset to Fig. 4). The DSRR structure is placed on an FR4 substrate of thickness $h_{FR4} = 1.51$ mm and area $23 \times 27\ mm^2$, while the thickness of the metallization is 0.035 mm. The real part of the permittivity constant of the substrate id $\varepsilon'_{FR4} = 4.34$, and the loss at around 2 GHz is $\tan(\delta) = 0.02$. An $SMV1405$ varactor diode was soldered into the DSRR, enabling separate dynamic tuning of the resonances at 2 and 3 kHz respectively by externally applied voltage. Diodes were used to implement the decoupling between the high frequency (RF) resonators and the low frequency biasing network. The setup was illuminated by a 2.88 GHz carrier (located at the far field of a horn antenna), polarized in the direction of the SRR's polarizability (Z-axis), and the backscattered fields was processed with a spectrum analyzer (PNA-L, Keysight). Both horizontal and vertical configurations, replicating the theoretical conditions discussed earlier, were considered. The resulting micro-Doppler combs were then collected for each distance between the scatterers for both configurations to produce Fig. 5.

Main spectral features of the coupled micro-Doppler combs, obtained theoretically, numerically and experimentally can be observed (Figs. 2, 3 and 5, respectively). A qualitative agreement can be seen between the figures, specifically the emergence of the fast decaying 1st, 5th and 7th peaks, which are the result of near field coupling. Again it should be noted that in the horizontal case, the distance starts from 1cm due to the finite size of the SRR's, which appears as a slower decay in comparison with the vertical case. The quantitative discrepancy between the figures is attributed to differences between the two fabricated samples, possessing slightly dissimilar resonances and their widths, in contrast with the simulated and theoretical plots. In order to co-locate resonances of the SRRs, an additional biasing voltage has been applied. However, this does not ensure the same resonance quality factors, which depend also on the imperfections in the fabrication. Furthermore, the capacitance of the varactor diodes does not depend linearly on the applied voltage, in contrast to the assumptions made in the theoretical and numerical investigations.

**Summary and Discussion**

Mixing phenomena between micro-Doppler combs, owing to the electromagnetic coupling between closely situated objects, have been studied theoretically, numerically and experimentally. Time-varying scattering properties of the targets were introduced parametrically into the electromagnetic problem and time-scales separation method has been applied. Time-varying coupled dipoles model is based on its well-known stationary counterpart, while the slow time modulation is accounted via expanding the expressions into Tailor series, where the ratio between the carrier and the modulation period is the small parameter. The numerical model is based on a series of frequency domain simulations, where all possible mutual configurations of internal impedances are considered (sampled). True time sequence of the events is then obtained by stitching the series of those instantaneous realizations and finally transforming back into the frequency domain to produce the coupled micro-Doppler comb at the baseband. The experimental realization in the GHz domain relies on spectral analysis of micro-Doppler combs at the baseband frequencies. Qualitative agreement between all the methods has been demonstrated, underlining the importance of electromagnetic coupling between closely situated targets.

The model problem, which was considered, is based on a pair of split ring resonators, controlled by a voltage drop on varactor diodes. However, similar behaviors are expected to emerge when different internal degrees of freedom within a target are involved. The most practically important ones include rotations and vibrations, which are almost always associated with motion of non-rigid bodies or complex gears. Here a simple analogy between a rotating wire and time-modulated dipole antenna has been demonstrated. It is worth noting that any periodic micro-Doppler signature can be emulated with the time modulation. For this purpose, the signal should be decomposed into a Fourier series and each of its coefficients should be fitted with a proper harmonic time modulation. This makes the developed emulation tool to be quite universal.

Micro-Doppler signatures, obtained from analysis of long-wavelength (in comparison to the object size) scattering problems, allows capturing information about a collective behavior of an object and its interaction with an environment. In more general context, micro-Doppler (or rotational Doppler) spectroscopy can be employed in stellar radiometry (e.g. for micro-turbulence investigations) in one extreme and for spinning molecules at the other [31]. Similar approaches can re-emerge in the case of nano-colloids, where large number of nano-scale particles drift, undergo Brownian motion and can demonstrate collective phenomena (nevertheless, different approaches to velocimetry can be employed, e.g. [32]). Examining this dynamics with laser beams and interferometry suggests using the tools, which are developed here, due to the face the long wavelength approximation is satisfied, allowing the investigation of collective dynamics.

**Acknowledgments**

The research was supported in part by Pazy Foundation and by ERC StG 'In Motion'.